\documentclass[showpacs,aps,preprintnumbers]{revtex4}
\usepackage{graphicx}
\usepackage{dcolumn}

\newcommand{\ep}{\epsilon}

\newcommand{\lraw}{\longrightarrow}

\newcommand{\pa}{\partial}

\newcommand{\td}{\tilde}
\newcommand{\sla}[1]{\slash\!\!\! #1}

\begin{document}

\title{The Scale on Chiral Symmetry Breaking}
\author{Mu-Lin Yan\footnote{mlyan@ustc.edu.cn}, Yi-Bin Huang\footnote{huangyb@mail.ustc.edu.cn}}

\affiliation{ CCAST(World Lab), P. O. Box 8730, Beijing, 100080, P. R. China \\
  and\\
Interdisciplinary Center for Theoretical Study, USTC, Hefei, Anhui
230026, P.R. China\footnote{mail address}}
\author{Xiao-Jun Wang}
\affiliation{Institute of Theoretical Physics, Beijing, 100080,
P.R. China}

\preprint{USTC-ICTS-03-7}

\begin{abstract}
We study the relation between the scale of chiral symmetry
spontaneously breaking and constituent quark mass. We argue that
this relation partly reveals strong interaction origination of
chiral symmetry breaking. We show that the relation can be
obtained via checking unitarity region of low-energy effective
field theory of QCD. This effective field theory must manifestly
include consistent quark mass as parameter. Thus we derive this
effective field theory from naive chiral constituent quark model.
The phenomenological value obtained by this method agree with
usual one determined by pion decay constant.
\end{abstract}


\maketitle

The typical difficulty on studies on QCD is from its dramatic
properties when dynamics of QCD lies in non-perturbative region.
The analysis of renormalization group shows that QCD is asymptotic
freedom at high energy scale, but should lie in confinement phase
at very low energy. Consequently, a phase transition must occur
when energy scale varies from higher to lower one. The phase
transition are dynamically characterized by well-known fermion
(quark) condensation phenomena. A dynamical scale (it is usually
referred as $\Lambda_{\rm QCD}$ in QCD) is consequently generated
by the quark condensation. It is just order parameter associating
the phase transition. Sometimes this scale is also transferred to
another effective parameter: so-called constituent quark mass $m$
and treated it as order parameter. Focusing on dynamics of QCD
with light flavor quarks only, however, the story is more
complicated: The quark condensation also breaks the (approximate)
global chiral symmetry of QCD. Accompanying with the chiral
symmetry spontaneously breaking (CSSB), another scale
$\Lambda_{\rm CSSB}$ must be dynamically generated and Goldstone
bosons appear as dynamical degrees of freedom. CSSB is one of the
most important features for the hadron physics. It together with
color confinement governs full low-energy dynamics of QCD. A
typical example is success of chiral perturbation theory
(ChPT)\cite{GL85a}.

An interesting issue is that global chiral symmetry is broken due
to pure strong interaction. To localize the global chiral symmetry
one has to introduce electroweak interactions. Because the quark
condensation breaks both of local and global chiral symmetry,
CSSB actually involves both of strong and electroweak interactions
(in contrast to color confinement caused by pure strong
interaction). The fact of electroweak relevance of CSSB has been
shown in determination on CSSB scale via weak decay constant of
pion \cite{MG84}, i.e., $\Lambda_{\rm CSSB}\sim 2\pi F_{\pi}\sim
1.2$GeV. The studies on role of strong interaction in CSSB,
however, seems to be more difficult, since complete understanding
on this issue requires underlying knowledge on dynamical mechanics
of color confinement. During the past decades, CSSB has been
extensively studied along this way, i.e., so called the formalism
of the gap equations (or Schwinger-Dyson equations, see
refs.\cite{Farhi,Roberts,Ripka,Hatsuda} and the references within,
and \cite{Cheng}). This method is rigorous and achieves some
successes, but still far from our final expectation so far.
Alternately, it should be also possible to explore CSSB by
starting within confinement phase. The key point is to find
relation between $\Lambda_{\rm CSSB}$ and constituent quark mass $m$ (or $\Lambda_{\rm QCD}$).
It will partly reflect the role of
strong interaction in CSSB. This is just purpose of this letter.
In this phase, the dynamical description is replaced by effective
one with constituent quarks and Goldstone bosons. In such effective
description a
critic-like energy scale must exist.  Above this energy scale, this
effective description on the
system collapses and below it, the description works. This critic-like
energy scale should
be just the scale of CSSB, $\Lambda_{\rm CSSB}$.

The natural criterion on whether an quantum effective field theory
(QEFT) description collapses or not is to check unitarity of the
QEFT. This claim bases on the fact that QEFT does not describe
full degrees of freedom of fundamental theory. When energy is
higher than characteristic scale of the QEFT, some new degrees of
freedoms will be excited consequently the unitarity of the QEFT is
lost. In this letter, we will derive a low-energy QEFT of pure
meson interaction from naive chiral constituent quark model, and
to obtain $\Lambda_{\rm CSSB}$ via checking unitarity region of
that QEFT.

It is well-known that a low energy effective meson theory should
be a well-defined perturbative theory in $N_c^{-1}$
expansion\cite{tH74}. Therefore, unitarity condition of
$S$-matrix, or optical theorem, has to satisfied order by order in
powers of $N_c^{-1}$ expansion,
\begin{eqnarray}\label{1}
{\rm Im}({\cal T}_{\beta,\alpha})_n=\frac{1}{2}\sum_{\rm
all\;\gamma}\sum_{m\leq n}({\cal T}_{\gamma,\alpha})_m
 ({\cal T}_{\gamma,\beta}^*)_{n-m}.
\end{eqnarray}
where the ${\cal T}_{\beta,\alpha}$ is transition amplitude from
state initial $\alpha$ to final state $\beta$, and $\gamma$
denotes all possible intermediate states on mass shells, and
${\cal T}_n\sim O((1/\sqrt{N_c})^n)$. According to standard power
counting law on large $N_c$ expansion in meson
interaction\cite{tH74,Dono90}, any transition amplitudes with
$n_V$ vertices, $n_e$ external meson lines, $n_i$ internal meson
lines and $n_l$ loops of mesons are of order
\begin{eqnarray}\label{2}
N_c^{n_V-n_i-n_e/2}=(N_c^{-\frac{1}{2}})^{2n_l+n_e-2},
\end{eqnarray}
where topological relation $n_l=n_i-n_V+1$ has been used. We focus
on transition amplitude from single meson initial state $\alpha$
to $k$ mesons final state
$\beta=\{\beta_1,\beta_2,\cdots,\beta_k\}$. Assuming intermediate
state $\gamma$ includes $s$ mesons
$\{\gamma_1,\gamma_2,\cdots,\gamma_s\}$, then using the power
counting rule (\ref{2}), eq. (\ref{1}) can be written as
\begin{eqnarray}\label{3}
{\rm Im}({\cal
T}_{\beta,\alpha})_{(2n_l+k-1)}=\frac{1}{2}\sum_{{\rm
all}\;\gamma(s)}\sum_{n'}
 ({\cal T}_{\gamma,\alpha})_{(2n'_l+s-1)}
 ({\cal T}_{\gamma,\beta}^*)_{(2n''_l+s+k-2)},
\end{eqnarray}
where $n_l$, $n'_l$ and $n''_l$ are meson loop numbers in
transition amplitude ${\cal T}_{\beta,\alpha}$, ${\cal
T}_{\gamma,\alpha}$ and ${\cal T}_{\gamma,\beta}$ respectively.
Both side of eq.~(\ref{3}) should be of the same order, thus
\begin{eqnarray}\label{4}
n'_l+n''_l+s=1+n_l.
\end{eqnarray}
For the case of leading order of transition amplitude ${\cal
T}_{\beta,\alpha}$, i.e., $n_l=0$, we have $n'_l=n''_l=0$ and
$s=1$ according to eq.~(\ref{4}). Consequently only
$\gamma=\alpha$ is allowed at the leading order. Since meson
fields are free point-particle at limit
$N_c\rightarrow\infty$\cite{tH74}, we have $({\cal
T}_{\alpha,\alpha})_0\equiv 0$. Therefore, it can be claimed that,
if any effective meson theory is unitary below its characteristic
scale, the on-shell transition amplitude from any meson state to
any multi-meson state must be real at leading order of $N_c^{-1}$
expansion,
\begin{eqnarray}\label{5}
{\rm Im}({\cal T}^{(0)}_{\beta,\alpha})_{k-1}=0.
\end{eqnarray}
where the superscript $(0)$ denotes the leading order of $N_c^{-1}$ expansion.
This claim will serve as equivalent description of unitarity for
any QEFTs on meson interaction.

A convenient effective description on the low energy QCD is naive
chiral constituent quark model (ChQM) proposed in ref.\cite{MG84}.
The constituent quark mass as order parameter associating to phase
transition is manifestly appear in this model. Thus this model
provides a possible framework to explore relation between
$\Lambda_{\rm CSSB}$ and order parameter. The simplest ChQM
is parameterized by the following SU(3)$_{V}$ invariant
Lagrangian
\begin{eqnarray}\label{6}
{\cal L}_{\rm ChQM}&=&i\bar{q}(\sla{\pa}+\sla{\Gamma}+
  g_{A}{\slash\!\!\!\!\Delta}\gamma_5-i\sla{V})q-m\bar{q}q
-\bar{q}Sq-\kappa\bar{q}P\gamma_5q \nonumber \\
&&+\frac{F^2}{16}<\nabla_\mu U\nabla^\mu U^{\dag}>
   +\frac{1}{4}m_0^2<V_\mu V^{\mu}>.
\end{eqnarray}
Here $V_\mu$ are vector meson octet, $<\cdots>$ denotes trace in
SU(3) flavor space, $\bar{q}=(\bar{u},\bar{d},\bar{s})$ are
constituent quark fields, $g_{A}=0.75$ is fitted by beta decay of
neutron, and
\begin{eqnarray}\label{7}
\Delta_\mu&=&\frac{1}{2}[\xi^{\dag}(\pa_\mu-ir_\mu)\xi
          -\xi(\pa_\mu-il_\mu)\xi^{\dag}], \nonumber \\
\Gamma_\mu&=&\frac{1}{2}[\xi^{\dag}(\pa_\mu-ir_\mu)\xi
          +\xi(\pa_\mu-il_\mu)\xi^{\dag}], \nonumber \\
\nabla_\mu U&=&\pa_\mu U-ir_\mu U+iUl_\mu=2\xi\Delta_\mu\xi,
   \\
\nabla_\mu U^{\dag}&=&\pa_\mu U^{\dag}-il_\mu
U^{\dag}+iU^{\dag}r_\mu
  =-2\xi^{\dag}\Delta_\mu\xi^{\dag}, \nonumber\\
S&=&\frac{1}{2}(\xi^{\dag}\td{\chi}\xi^{\dag}+\xi\td{\chi}^{\dag}\xi),
\hspace{1in}
P=\frac{1}{2}(\xi^{\dag}\td{\chi}\xi^{\dag}-\xi\td{\chi}^{\dag}\xi),
\nonumber
\end{eqnarray}
where $l_\mu=v_\mu+a_\mu$ and $r_\mu=v_\mu-a_\mu$,
$\td{\chi}=s_{\rm ext}+{\cal M}+ip$ with external fields $v_\mu$
(vector), $a_\mu$ (axial-vector), $s_{\rm ext}$ (scalar), $p$
(pseudoscalar), and current quark mass matrix ${\cal M}={\rm
diag}\{m_u,m_d,m_s\}$ respectively. $\xi$ associates with
non-linear realization of spontaneously broken global chiral
symmetry $G=SU(3)_L\times SU(3)_R$ introduced by Weinberg
\cite{Wein68},
\begin{equation}\label{8}
\xi(\Phi)\rightarrow
g_R\xi(\Phi)h^{\dag}(\Phi)=h(\Phi)\xi(\Phi)g_L^{\dag},\hspace{0.5in}
 g_L, g_R\in G,\;\;h(\Phi)\in H=SU(3)_{V}.
\end{equation}
Explicit form of $\xi(\Phi)$ is usually taken as
\begin{equation}\label{9}
\xi(\Phi)=\exp{\{i\lambda^a \Phi^a(x)/2\}},\hspace{1in}
U(\Phi)=\xi^2(\Phi),
\end{equation}
where $\lambda^1,\cdots,\lambda^8$ are SU(3) Gell-Mann matrices in
flavor space, and the Goldstone bosons $\Phi^a$ are identified to
pseudoscalar meson octet.

The transformation law under SU(3)$_{V}$ for any quantities
defined in eqs.~(\ref{6}) and (\ref{8}) are
\begin{eqnarray}\label{10}
  q&\lraw & h(\Phi)q, \hspace{0.6in}
\Delta_\mu\lraw h(\Phi)\Delta_\mu h^{\dag}(\Phi), \hspace{0.6in}
V_\mu\rightarrow h(\Phi)V_\mu h^{\dag}(\Phi), \nonumber \\
\Gamma_\mu & \lraw & h(\Phi)\Gamma_\mu
h^{\dag}(\Phi)+h(\Phi)\pa_\mu
  h^{\dag}(\Phi).
\end{eqnarray}
The homogenous transformation law on vector meson field is usually
referred as WCCWZ realization on vector meson.\cite{Wein68,WCCWZ}

Because there is no kinetic term for vector fields in ${\cal
L}_{\rm ChQM}$, they serve as auxiliary fields in this formalism.
From the equation of motion $\delta{\cal L}_{\rm ChQM}/\delta
V_{\mu}=0$, we can see the vector fields in ${\cal L}_{\rm ChQM}$
are the composite fields of constituent quarks, Therefore, WCCWZ
method is actually a way to catch the effects of constituent quark
bound states in the ChQM. $F,\;g_{A},\;m,\;\kappa$ and $m_0$ in
eq.~(\ref{6}) are free parameters of the model.

The effective action on meson interaction, $S_{\rm eff}[U,V]$, can
be obtained via integrating out quark fields,
\begin{eqnarray}\label{11}
S_{\rm eff}[U,V]=\ln \det ({\cal D})+\int d^4x
\{\frac{F^2}{16}<\nabla_\mu U\nabla^\mu U^{\dag}>
   +\frac{1}{4}m_0^2<V_\mu V^{\mu}>\},
\end{eqnarray}
where ${\cal D}=\sla{\pa}+\sla{\Gamma}+
  g_{A}{\slash\!\!\!\!\Delta}\gamma_5-i\sla{V}-m
-S-\kappa P\gamma_5$, $F$ and $m_0$ will receive quark loop
effects and then are renormalized into $F_\pi=186{\rm MeV}$ and
the physical masses $m_V$ of vector mesons respectively. Then
$S_{\rm eff}[U,V]$ parameterizes an QEFT on pure meson interaction.

Now let us consider unitarity of this QEFT. In particular, we
focus on $V\to\Phi\Phi$ decay amplitude and impose eq.~(\ref{5})
to find unitarity region of the QEFT. To separate relevant
effective action from $S_{\rm eff}[U,V]$ and rewrite it into
appropriate form
\begin{eqnarray}\label{12}
S_{\rm eff}^{V\Phi\Phi}= \sum\limits_{abc}\int\frac{d^4p d^4q_1
d^4q_2}{(2\pi2\pi)^4}\delta(p+q_1+q_2)
V_\mu^{ab}(p)\Phi^{bc}(q_1)\Phi^{ca}(q_2)q_2^\mu
f_{abc}(p^2,q_1^2,q_2^2),
\end{eqnarray}
we have
\begin{eqnarray}\label{13}
{\cal T}_{\Phi\Phi, V}^{(0)}\equiv
<\Phi^{bc}(q_1)\Phi^{ca}(q_2)|{\cal
T}^{(0)}|V^{ab}(p,\lambda)>=(2\pi)^4\delta^4(p-q_1-q_2)q_2^\mu
\epsilon_\mu^\lambda f_{abc}(p^2,q_1^2,q_2^2).
\end{eqnarray}
where $\epsilon_\mu^\lambda $ is the polarization vector of the
vector meson $V^{ab}(p,\lambda)$. Consequently
\begin{eqnarray}\label{14}
{\rm Im}{\cal T}^{(0)}_{\Phi\Phi,V}\propto {\rm Im}
f_{abc}(p^2,q_1^2,q_2^2).
\end{eqnarray}
The form factor $f_{abc}(p^2,q_1^2,q_2^2)$ can be rewritten as
$f_{abc}(p^2,q_1^2,q_2^2)=f_2(p^2)+f_3(p^2,q_1^2,q_2^2)$, where
$f_2$ and $f_3$, with subscript $abc$ suppressed, are two-point
Green function (Fig. 1-a) and three-point Green function (Fig.1-b)
of constituent quark fields respectively, and are linearly
independent.
\begin{figure}[pthb]
   \centering
   \includegraphics[width=5in]{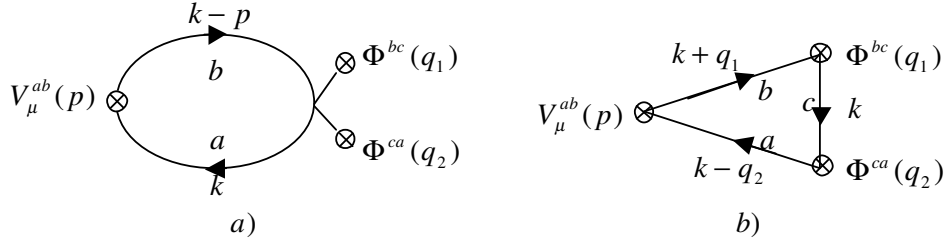}
\begin{minipage}{5in}
   \caption{Two-point and three-point diagrams of quark loops for
   effective action $S^{V\Phi\Phi}$.}
\end{minipage}
\end{figure}
Explicitly, the calculations on form factors $f_2(p^2)$ and
$f_3(p^2,q_1^2,q_2^2)$ are straightforward,
\begin{eqnarray}\label{15}
f_2(p^2)&=&
i\int\frac{d^4k}{(2\pi)^4}\frac{g(k,p)}{[(k-p)^2-M_b^2+i\ep]
(k^2-M_a^2+i\ep)} =\int_0^1 dx\int_0^\infty du\frac{{\td
g}(x,u,p^2)}{(u+D_2-i\ep)^2},
 \nonumber \\
f_3(p^2,q_1^2,q_2^2)&=&
i\int\frac{d^4k}{(2\pi)^4}\frac{h(k,p,q_1,q_2)}{[(k+q_1)^2-M_b^2+i\ep]
(k^2-M_c^2+i\ep)[(k-q_2)^2-M_a^2+i\ep]}\nonumber \\
&=&\int_0^1 xdx\int_0^1 dy\int_0^\infty du\frac{{\td
h}(x,y,u,p^2,q_1^2,q_2^2)}{(u+D_3-i\ep)^3},
\end{eqnarray}
where $g,\;h$ and ${\td g},\; \td{h}$ are definite real and
polynomial functions of $u=k_E^2$ ($k_E^\mu=-ik_0,k_x,k_y,k_z$),
$M_{a}=m+m_{a}\;\;(a=u,d,s)$ and
\begin{eqnarray}\label{16}
D_2&=&M_a^2(1-x)+M_b^2x-p^2x(1-x), \nonumber \\
D_3&=&M_a^2x(1-y)+M_b^2(1-x)+M_c^2xy-p^2x(1-x)(1-y)
-q_1^2xy(1-x)-q_2^2x^2y(1-y).
\end{eqnarray}
Using principle value formula
\begin{eqnarray}\label{17}
\frac{1}{z\pm i\ep}=\frac{{\cal P}}{z}\mp i\pi\delta(z),
\end{eqnarray}
where $\frac{{\cal P}}{z}$ is the principle value and is real, we
can express $f_2$ and $f_3$ as
\begin{eqnarray}\label{18}
f_2(p^2)&=&-\int_0^1 dx\int_0^\infty du{\td g}\frac{\pa}{\pa
u}\Big(\frac{{\cal P}}{u+D_1}+i\pi\delta(u+D_1)\Big), \nonumber \\
f_3(p^2,q_1^2,q_2^2)&=&\frac{1}{2}\int_0^1 xdx\int_0^1
dy\int_0^\infty du{\td h}\frac{\pa^2}{\pa u^2}\Big(\frac{{\cal
P}}{u+D_2}+i\pi\delta(u+D_2)\Big).
\end{eqnarray}
Then we obtain
\begin{eqnarray}\label{19}
{\rm Im}f_2(p^2)&\propto &\int_0^1 dx\int_0^\infty du\frac{\pa
{\td g(x,u,p^2)}}{\pa u}\delta(u+D_2), \nonumber \\
{\rm Im}f_3(p^2,q_1^2,q_2^2)&\propto &\int_0^1 xdx\int_0^1
dy\int_0^\infty du\frac{\pa^2{\td h(x,y,u,p^2,q_1^2,q_2^2)}}{\pa
u^2}\delta(u+D_3).
\end{eqnarray}
Finally we have
\begin{eqnarray}\label{20}
{\rm Im}f_i(p^2)=0\Longleftrightarrow
u+D_i\neq0\Longleftrightarrow D_i>0\hspace{1in}i=2,3,
\end{eqnarray}
where $u>0,0<x,y<1$ have been considered.
More precisely,
\begin{eqnarray}\label{21}
{\rm Im}{\cal T}_{\Phi\Phi,V}^{(0)}=0\Longleftrightarrow
\left\{\begin{array}{cc}D_1>0&(0<x<1)\\
D_2>0&\;\;\;\;(0<x,y<1).\end{array}\right.
\end{eqnarray}
This will lead to a restriction on the range of $p^2$. The former
inequality will hold in domain $0\leq x\leq1$ if and only if
$$
M_{V^{ab}}=\sqrt{p^2}\leq M_a+M_b.
$$
As to the latter, the right side of it has no stationary point in
$x-y$ plane, therefore this inequality holding in the square
domain is equivalent to it holding at boundary of the square,
which gives
$$
\left\{\begin{array}{c}
 M_{V^{ab}}=\sqrt{p^2}\leq M_a+M_b,\\
 M_{\Phi^{ab}}=\sqrt{q^2}\leq M_a+M_b.
\end{array}
\right.
$$
Because $M_{\Phi^{ab}}<M_{V^{ab}}$, we see that the second
condition is satisfied if the first one does. Therefore we
conclude that the necessary condition for the effective theory to
be unitary is
\begin{eqnarray}\label{22}
M_{V^{ab}}=\sqrt{p^2}\leq \Lambda^{ab}\equiv 2m+m_a+m_b.
\end{eqnarray}

In the beginning of this letter, we have actually argued an
important fact that $\Lambda^{ab} \equiv 2m+m_a+m_b$ is a
critical energy scale in the meson QEFT parameterized by $S_{\rm
eff}[U,V]$. As $\sqrt{p^2}$ is below $ \Lambda^{ab}$, the
$S$-matrices yielded from the Feynman rules of that QEFT are
unitary, while as $\sqrt{p^2}$ is above this scale, the unitarity
of that QEFT will be violated. This fact indicates that the
well-defined QEFT describing the meson physics in the framework
of ChQM exists only as the characteristic energy is below
$\Lambda^{ab}$. When energy is above $\Lambda^{ab}$, the
effective meson Lagrangian description of the dynamics is illegal
in principle because the unitarity fails. This is precisely a
critical phenomenon, or quantum phase transition in quantum field
theory, which is caused by quantum fluctuations in the
system\cite{Sachdev}. Recalling the meaning of the scale
$\Lambda_{\rm CSSB}$ of chiral symmetry spontaneously breaking in
QCD, we can see that $\Lambda^{ab}$ play the same role as
$\Lambda_{\rm CSSB}$. Then, in the framework of ChQM, we identify
\begin{equation}\label{23}
\Lambda_{\rm CSSB}=\Lambda^{ab}\equiv 2m+m_a+m_b.
\end{equation}

The above equation just explores a simple relation between
$\Lambda_{\rm CSSB}$ and constituent quark mass, and is the main result of this letter.
It should be
notice that $\Lambda_{\rm CSSB}$ is flavor-dependent as we
expect. This reflects the fact that although the vacuum (quark
condensation) is $SU(3)_{\rm V}$ invariant at chiral limit, it is
explicitly broken to Abelian subgroup of $SU(3)_{\rm V}$ when
current quark masses are turned on. As discussed at the beginning of
this letter, $\Lambda_{\rm QCD}$ should be unique scale of QCD at
low energy. In other words, other dimensional quantities, even
including $\Lambda_{\rm CSSB}$, should be related to
$\Lambda_{\rm QCD}$. Thus more fundamental task is to find
relation between $\Lambda_{\rm CSSB}$ and $\Lambda_{\rm QCD}$
from eq.~(\ref{23}). Roughly we can expect $m\simeq\Lambda_{\rm
QCD}$, at some definite low energy limits at least. However, the
precise coefficient is no longer 1. A direct evidence is that if
interaction between gluons and constituent quarks are turned on
(this coupling is usually expected to be weak, but should not
vanish exactly), the self-energy diagram of constituent quarks
will contribute to mass term of constituent quarks. To explore
relation between $\Lambda_{\rm CSSB}$ and $\Lambda_{\rm QCD}$
means that we should explore exactly relation between
$\Lambda_{\rm QCD}$ and constituent quark mass. It actually
requires that we should know underlying dynamical mechanism of low
energy QCD and thus will be great challenge.

Phenomenologically, it is also interesting to fix numerical value
of $\Lambda_{\rm CSSB}$. The low-energy limit of the QEFT can
obtained via integrating out vector meson
fields\cite{Wang98,Eck89}. It means that, at very low energy, the
dynamics of vector mesons are replaced by pseudoscalar meson
fields. Expanding the resulted Lagrangian up to ${\cal O}(p^4)$
in terms of Schwinger's proper time method\cite{Sch54,Ball89}, we
get ${\cal O}(p^4)$ ChPT-coefficients as follows
\begin{eqnarray}\label{24}
L_1&=&\frac{1}{2}L_2=\frac{1}{128\pi^2}, \hspace{0.8in}
L_3=-\frac{3}{64\pi^2}+\frac{1}{64\pi^2}g_A^4, \nonumber \\
L_4&=&L_6=0, \hspace{1.2in}L_5=\frac{3m}{32\pi^2B_0}g_A^2,\nonumber \\
L_8&=&\frac{F_\pi^2}{128B_0m}(3-\kappa^2)+\frac{3m}{64\pi^2B_0}
  (\frac{m}{B_0}-\kappa g_A-\frac{g_A^2}{2}-\frac{B_0}{6m}g_A^2)
  +\frac{L_5}{2},\nonumber \\
L_9&=&\frac{1}{16\pi^2}, \hspace{1.35in}
L_{10}=-\frac{1}{16\pi^2}+\frac{1}{32\pi^2}g_A^2.
\end{eqnarray}
The above expressions of $L_i$ have been obtained in some previous
refs.\cite{Esp90,Wang98,Bijnens93} (except $L_8$). Then inputting
experimental values of $L_5$ and $L_8$ and takeing $g_{_A}=0.75$
(fitted by $n\rightarrow pe^-\bar{\nu}_e$ decay\cite{MG84}) and
$m_u+m_d\simeq 11$MeV, we can fix phenomenological values of other
free parameters as $B_0\simeq 1.8$GeV, $m\simeq 460$MeV and
$\kappa\simeq0.5$. The numerical results for those low energy
constants are listed in table II.

\begin{table}[pht]
\centering
 \begin{tabular}{cccccccccc}\hline \\
&$L_1$&$L_2$&$L_3$&$L_4$&$L_5$&$L_6$&$L_8$&$L_9$&$L_{10}$
  \\ \hline
ChPT&$0.7\pm 0.3$&$1.3\pm 0.7$&$-4.4\pm 2.5$&$-0.3\pm 0.5$&$1.4\pm
  0.5$&$-0.2\pm 0.3$&$0.9\pm 0.3$&$6.9\pm 0.7$&$-5.2\pm 0.3$\\
{ChQM}&0.79&1.58&-4.25&0&$1.4^{a)}$&0&$0.9^{a)}$&6.33&-4.55 \\
\hline
   \end{tabular}
\begin{minipage}{6in}
\caption{\small $L_i$ in units of $10^{-3}$, ${\mu}=m_\rho$.
a)input. b)contribution from gluon anomaly.}
\end{minipage}
\end{table}

Numerically, for $ud$-flavor system (e.g., $\pi-\rho-\omega$
physics),
\begin{equation}\label{ud}
\Lambda_{\rm CSSB}(ud)\simeq 2m=920{\rm MeV}.
\end{equation}
For $u(d)s$-flavor system (e.g., $K-K^*$ physics),
\begin{equation}\label{ud-s}
\Lambda_{\rm CSSB}(u(d)s)\simeq 2m+m_s=1090{\rm MeV}.
\end{equation}
For $\bar{s}s$ case (e.g., $\phi$-physics),
\begin{equation}\label{ss}
\Lambda_{\rm CSSB}(ss)\simeq 2(m+m_s)=1260{\rm MeV}.
\end{equation}
Since $m_\rho<\Lambda_{\rm CSSB}(ud)$, $m_{K^*}<\Lambda_{\rm
CSSB}(u(d)s)$ and $m_\phi<\Lambda_{\rm CSSB}(ss)$, the effective
meson field theory derived by resummation derivation in ChQM in
this paper is unitary. And the low energy expansions in powers of
$p$ are legitimate and convergent due to $p^2/\Lambda_{\rm
CSSB}^2<1$. It means that all light flavor vector meson resonances
can be included in ChQM consistently. It is remarkable that the
quantum phase transitions in ChQM can be explored successfully in
resummation derivation method, and the corresponding critical
scales are determined analytically.

To conclude, it is shown that the scale of CSSB is not independent
of the scale of color confinement. The relation between two scales
reveals strong interaction origination of CSSB phenomena.
However, to explore this relation precisely is very difficult due
to lack of underlying knowledge on color confinement. Instead we
argued that this relation can be partly replaced by one between
$\Lambda_{\rm CSSB}$ and constituent quark mass $m$. We used
naive chiral constituent quark model to find this relation via
checking unitarity region of induced QEFT of meson interaction.
Phenomenologically, we determined numerical value of
$\Lambda_{\rm CSSB}$ in terms of consistent fit on values of free
parameters of ChQM. The result agree with usual value of
$\Lambda_{\rm CSSB}$ determined by pion decay constant. Our
evaluation also shows that lowest order vector meson resonances
can be consistently included in naive ChQM.

\begin{center}
{\bf ACKNOWLEDGMENTS}
\end{center}
This work is partially supported by NSF of China 90103002 and the
Grant of the Chinese Academy of Sciences. The authors wish to
thank Yong-Shi Wu (Utah U) for his stimulating discussions on
quantum phase transitions.


\begin{thebibliography}{99}
\bibitem{GL85a}J.Gasser and H.Leutwyler, Ann. Phys. {\bf 158}(1984) 142;
   Nucl. Phys. {\bf B250}(1985) 465.
\bibitem{MG84}A. Manohar and H. Georgi, Nucl. Phys. {\bf B234} (1984)
    189; H. Georgi, {\it Weak Interactions and Modern Particle Theory}
    (Benjamin/Cimmings, Menlo Park, CA, 1984) sect. 6.
\bibitem{Farhi} {\it Dynamical Gauge Symmetry Breaking}, ed.
E.Farhi and R.Jackiw, Word Scientific, (1982); {\it Proceedings of
the 1991 Nagoya Spring School on Dynamical Symmetry Breaking,} ed.
K.Yamawaki, Word Scientific, Singapore, (1992).
\bibitem{Roberts} C. D. Roberts, nucl-th/0007054.
\bibitem{Ripka} G. Ripka, {\it Quarks Bound by Chiral Fields}, Clarendon Press,
Oxford, (1997).
\bibitem{Hatsuda} T. Hatsuda and T. Kunihiro, Phys. Rep. {\bf 247}
(1994) 221.
\bibitem{Cheng}G. Cheng and T.K.Kuo, J. Math Phys. {\bf 38}, (1997) 6119.
\bibitem{tH74}G. 't Hooft, Nucl. Phys. {\bf B75} (1974) 461.
\bibitem{Dono90}J.F. Donoghue, E.Golowich and B.R.Holstein, {\it Dynamics of the Standard Model},
pp258-272, Cambridge Univ Press, (1992).
\bibitem{Wein68}S. Weinberg, Phys. Rev. {\bf 166} (1968) 1568.
\bibitem{WCCWZ}S. Coleman, J. Wess and B. Zumino, Phys. Rev. {\bf 177}
(1969) 2239;C. G. Callan, S. Coleman, J. Wess and B. Zumino, {\it
ibid} 2247.
\bibitem{Sachdev} e.g., see, S. Sachdev, {\it Quantum Phase Transitions},
Cambridge Univ. Press, 1999.
\bibitem{Wang98}X. J. Wang and M. L. Yan, Jour. Phys. {\bf G24} (1998)
1077.
\bibitem{Eck89}G. Ecker, J.Gasser, A. Pich and E.de Rafel, Nucl. Phys.
{\bf B321} (1989) 311; G. Ecker, H. Leutwyer, J. Gasser, A. Pich
and E.de Rafel, Phys. Lett. {\bf B223} (1989) 425.
\bibitem{Sch54}J. Schwinger, Phys. Rev. {\bf 82} (1951) 664; J. Schwinger,
Phys, Rev. {\bf 93} (1954) 615.
\bibitem{Ball89}R. D. Ball, Phys. Rep. {\bf 182} (1989) 1.
\bibitem{Li95} B. A. Li, Phys. Rev. {\bf D52} (1995) 5165, 5184.
\bibitem{Esp90}D. Espriu, E. de Rafael and J. Taron, Nucl. Phys. {\bf
B345} (1990) 22.
\bibitem{Bijnens93} J. Bijnens, C. Bruno and E. de Rafael, Nucl. Phys. {\bf
B390} (1993) 501.
\end{thebibliography}
\end{document}